\documentclass[]{aa}
\usepackage{epsfig}
\begin{document}
\def\lsim{\vcenter{\hbox{$<$}\offinterlineskip\hbox{$\sim$}}}
\def\gsim{\vcenter{\hbox{$>$}\offinterlineskip\hbox{$\sim$}}}
\thesaurus{06(08.02.4 ; 08.03.4; 08.09.2; 08.13.2; 08.16.4; 08.19.3)}
\title{The circumstellar envelope of AFGL 4106\thanks{based on observations
       obtained at the European Southern Observatory, La Silla, Chile}}
\author{Jacco Th. van Loon\inst{1}, F.J. Molster\inst{1}, Hans van
        Winckel\inst{2} \and L.B.F.M. Waters\inst{1,3}}
\institute{Astronomical Institute, University of Amsterdam, Kruislaan 403,
           NL-1098 SJ Amsterdam, The Netherlands
      \and Institute of Astronomy, Catholic University of Leuven,
           Celestijnenlaan 200B, B-3001 Heverlee, Belgium
      \and Space Research Organization Netherlands, Landleven 12, NL-9700 AV
           Groningen, The Netherlands}
\offprints{Jacco Th.\ van Loon, jacco@ast.cam.ac.uk}
\date{Received date; accepted date}
\maketitle
\markboth{Jacco Th.\ van Loon et al.: The circumstellar envelope of AFGL 4106}
         {Jacco Th.\ van Loon et al.: The circumstellar envelope of AFGL 4106}
\begin{abstract}

We present new imaging and spectroscopy of the post-red supergiant binary AFGL
4106. Coronographic imaging in H$\alpha$ reveals the shape and extent of the
ionized region in the circumstellar envelope (CSE). Echelle spectroscopy with
the slit covering almost the entire extent of the CSE is used to derive the
physical conditions in the ionized region and the optical depth of the dust
contained within the CSE.

The dust shell around AFGL 4106 is clumpy and mixed with ionized gas.
H$\alpha$ and [N~{\sc ii}] emission is brightest from a thin bow-shaped layer
just outside of the detached dust shell. On-going mass loss is traced by
[Ca~{\sc ii}] emission and blue-shifted absorption in lines of low-ionization
species. A simple model is used to interpret the spatial distribution of the
circumstellar extinction and the dust emission in a consistent way.

\keywords{binaries: spectroscopic --- circumstellar matter --- Stars:
individual: AFGL 4106 --- Stars: mass loss --- Stars: AGB and post-AGB ---
supergiants}
\end{abstract}

\section{Introduction}

In their final stages of evolution, both massive ($M{\gsim}8$ M$_\odot$) and
intermediate-mass ($1{\lsim}M{\lsim}8$ M$_\odot$) stars can become
dust-enshrouded as a result of intense mass loss at rates of $10^{-5}$ up to
$10^{-3}$ M$_\odot$ yr$^{-1}$. For the massive stars, this heavy mass loss
occurs during the red supergiant (RSG) phase, while for the intermediate-mass
stars this happens during the Asymptotic Giant Branch (AGB) evolution. After
having lost a significant fraction of their initial mass during this episode,
the photospheric temperature of the star increases again and the stellar
ultraviolet (UV) radiation field starts ionizing the dusty CSE, producing
either a Planetary Nebula (PN) or a post-RSG nebula surrounding a Wolf-Rayet
star. The mechanisms at play during the short-lasting transition stage between
RSG/AGB and WR/PN are still poorly known. Especially the final evolution of
the massive objects is hitherto poorly documented. One such massive object is
AFGL 4106.

Garc\'{\i}a-Lario et al.\ (1994) discovered nebular line emission from the IR
object AFGL 4106, which came as a surprise because the central star was
thought to be of spectral type G: too cool to ionize a CSE. The dust shell
around AFGL 4106 was imaged in the infrared (IR) by Molster et al.\ (1999) who
modelled the spectral energy distribution as observed by ISO with a radiation
transfer code. They provide evidence that the object is a binary consisting of
a late-A/early-F type star and a somewhat fainter M-type companion. Their
distance estimate yields luminosities too high for AGB evolution and hence
AFGL 4106 must be the result of post-RSG evolution.

Here we present and analyse new observations of this object and focus on the
spatial distribution and physical conditions of the CSE around AFGL4106.
Coronographic imaging in H$\alpha$ shows the spatial distribution of the
ionized material around AFGL 4106. Spatially resolved echelle spectroscopy is
performed to measure expansion velocities, electron densities and the internal
extinction by the dust. The spatial distribution of dust extinction and
emission is modelled in a consistent way. Finally we discuss the structure of
the circumstellar envelope and the recent mass-loss history of AFGL 4106.

\section{Observations}

\subsection{H$\alpha$ coronographic imaging}

On February 4, 1996, we used the multi-mode instrument EMMI at the ESO 3.5m
NTT on La Silla, Chile, to image AFGL 4106 through a filter centred at
H$\alpha$, with a rather narrow width of 33 \AA\ to avoid contamination by
[N~{\sc ii}] emission. Because of the high apparent brightness of the star
(9$^{th}$ mag in V) coronographic techniques were used to limit the saturation
of the CCD. A glass plate was inserted in the aperture wheel, holding six dark
blots of different size and attenuation. The star was placed behind a
1.9$^{\prime\prime}$ blot with a central attenuation of a few mag. Two frames
of each 5 min integration time in H$\alpha$ were combined. A 10 s exposure
through a broad-band Johnson R filter was used to correct for the continuum
contribution within the H$\alpha$ filter band. The R-band image was aligned
with the H$\alpha$ image and scaled linearly by comparing the intensities of
several field stars. Before subtracting the R-band image from the H$\alpha$
image both were corrected for structure in the instrumental response over the
pixels of the CCD by dividing them by images of the morning twilight sky,
taken through the corresponding filters. Remaining CCD artifacts and cosmic
ray impacts were removed by hand. The seeing was constant at
1$^{\prime\prime}$, and the pixel size was 0.268$^{\prime\prime}$.

\subsection{Echelle spectroscopic imaging}

On February 4, 1996, we used the same telescope and instrument to take echelle
spectra of AFGL 4106 with several slit orientations. The wavelength region
extended from 6000 to 8350 \AA. Using a slit width of 1$^{\prime\prime}$ a
resolving power of $R\sim7.5\times10^4$ was obtained corresponding to a
velocity resolution of 4 km s$^{-1}$. The slit length was set to
15$^{\prime\prime}$ and the spatial resolution was 0.268$^{\prime\prime}$ per
pixel. We centred the star in the middle of the slit but obtained 15 min
spectra with different position angles: 0$^{\circ}$, 45$^{\circ}$,
90$^{\circ}$ and 135$^{\circ}$ where the position angle is defined from West
over North. The reduction of the spectra was standard and included
flat-fielding, wavelength calibration on the basis of Th-Ar exposures,
spectral response correction on the basis of a measured spectrum of standard
star HD 60753, interactive cosmic hit cleaning, extinction correction with
average extinction values, and finally absolute flux calibration. The seeing
was about 0.8 to 0.9$^{\prime\prime}$.

\section{Results}

\subsection{H$\alpha$ coronographic imaging}

%
%
\begin{figure}[tb]
\centerline{\psfig{figure=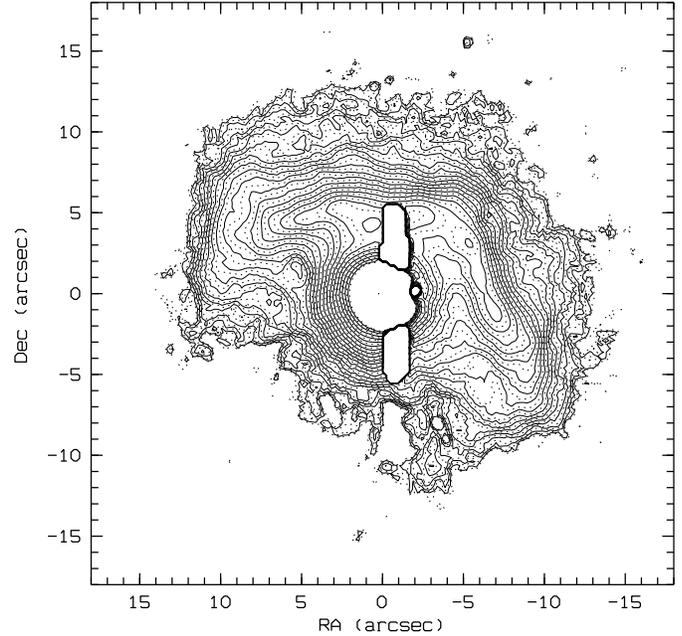,width=88mm}}
\caption[]{Coronographic image in the light of H$\alpha$ on a logarithmic flux
density scale. The flux density level of the brightest part of the extended
emission is $\sim7\times10^{-15}$ W m$^{-2} \mu$m$^{-1}$ arcsec$^{-2}$. The
faintest emission visible in the picture is a factor of ten fainter. North is
up, and East is to the left.}
\end{figure}

The coronographic image in H$\alpha$ is displayed on a logarithmic flux
density scale in Fig.\ 1. North is up, and East is to the left. The star
itself is heavily saturated despite the use of the coronographic blot,
resulting in the artificial NS-orientated bar. The absolute flux density
calibration was derived from the spectra and is estimated to be accurate
within $\sim20$\%. The flux density level of the brightest part of the
extended emission is $\sim7\times10^{-15}$ W m$^{-2} \mu$m$^{-1}$
arcsec$^{-2}$. The faintest emission visible in the picture is about a factor
of ten fainter.

AFGL 4106 appears to be situated in a bright, spatially extended emission
complex. From N to W of the star, the strongest emission delineates a
bow-shaped structure much akin the bow-shocks associated with stars that move
supersonically through the interstellar medium (Kaper et al.\ 1997). The
emission bow is located at a projected distance of 5$^{\prime\prime}$ up to
10$^{\prime\prime}$ from the central star. The faintest H$\alpha$ emission is
detected up to $\sim14^{\prime\prime}$ from the central star. In the SE much
less H$\alpha$ emission is detected. Interestingly, the 10 $\mu$m emission
(Molster et al.\ 1999) shows a clear anti-correlation with the H$\alpha$
emission: the 10 $\mu$m emission peaks in the SE and is faintest in the NW.

\subsection{Echelle spectroscopic imaging}

\subsubsection{Absorption lines}

The spectral type has been redetermined by van Winckel et al.\ (in
preparation) and Molster et al.\ (1999). The optical spectrum is dominated by
a late-A or early-F type star (T$_{\rm eff}\sim7500$ K) with a
nitrogen-enhanced photosphere resulting in strong N~{\sc i} absorption lines
around 8000 \AA. They also spectroscopically discovered the presence of an
early-M type companion star (T$_{\rm eff}\sim3750$ K). The self-absorbed
H$\alpha$ line profile centred on the star is identical to that observed by
Garc\'{\i}a-Lario et al.\ (1994). This suggests that the H$\alpha$ emission
conditions in the vicinity of the star have been stable over a period of at
least six years.

%
%
\begin{figure}[tb]
\centerline{\psfig{figure=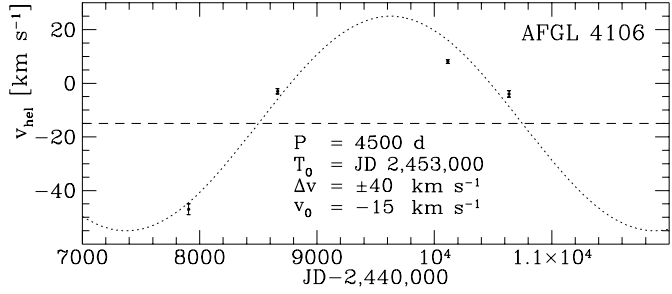,width=88mm}}
\caption[]{Radial velocity curve for AFGL 4106. The solution is not unique,
but periods much longer than 12 yr are difficult to reconcile with this data.}
\end{figure}

A total of 31 absorption lines of the atomic and singly ionized species
Fe~{\sc i}, Fe~{\sc ii}, Sc~{\sc ii}, Si~{\sc ii}, O~{\sc i}, Mg~{\sc ii},
Na~{\sc i} and N~{\sc i} were selected for measuring the star's radial
velocity. Considering a heliocentric correction for the movement of the Earth
of $+13.3$ km s$^{-1}$, the heliocentric velocity of the star is determined at
$v_\star=+8.1\pm0.7$ km s$^{-1}$. For an individual line, the typical
deviation from the mean was 4 km s$^{-1}$, which is equal to the spectral
resolution. The stellar velocity differs considerably from the value of
$v_\star=-47\pm2$ km s$^{-1}$ derived by Garc\'{\i}a-Lario et al.\ (1994) from
the Si~{\sc ii} $\lambda\lambda$6347,6371 lines (after transforming from LSR
to heliocentre) in their spectra taken in February 1990. In March 1992,
$^{12}$CO J=1$\rightarrow$0 emission at 115 GHz was detected by Josselin et
al.\ (1998). From the mean of the velocities of the blue- and red-most CO
emission we estimate $v_\star=-3\pm1$ km s$^{-1}$. An additional measurement
from photospheric absorption lines in high resolution spectra taken in July
1996 (van Winckel et al., in preparation) yields $v_\star=-4\pm1.2$ km
s$^{-1}$. An approximate radial velocity curve (Fig.\ 2) for the maximum
period possible, assuming circular orbits, yields $P_{\rm max}=4500$ d. For
(currently) equal masses of the two stars Kepler's third law yields the mass
of each star: $M\sim15\sin^{-3}i$ M$_\odot$ with orbital inclination angle $i$
(shorter periods imply smaller masses).

%
%
\begin{figure}[tb]
\centerline{\psfig{figure=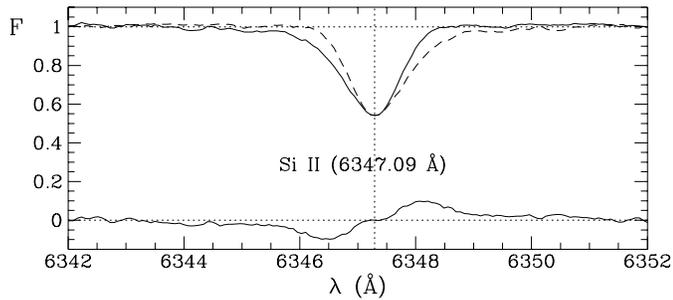,width=88mm}}
\caption[]{Spectrum around Si~{\sc ii} $\lambda$6347 \AA\ (solid) and the same
spectrum mirrored with respect to 6347.29 \AA\ where the line profile is
deepest (dashed). Below, their difference spectrum is plotted. The wavelength
axis is heliocentric.}
\end{figure}

The Si~{\sc ii} $\lambda\lambda$6347,6371, Fe~{\sc ii} $\lambda$6456, N~{\sc
i} $\lambda$7468 and Fe~{\sc i} $\lambda$8327 absorption lines have asymmetric
line profiles, with more absorption in the blue wing. This may be caused by
outflowing matter due to present day mass loss. The spectrum around Si~{\sc
ii} $\lambda$6347 is shown as an example (Fig.\ 3). The spectrum was
normalised by dividing by a constant equal to the spectrum level outside of
the absorption line, i.e.\ the spectral slope is not affected. Also plotted is
the difference between this spectrum, and the spectrum after mirroring with
respect to the absorption extremum at 6347.29 \AA (after heliocentric
correction). This difference spectrum shows the absorption excess in the blue
wing of the line profile, reaching a maximum at a blueshift of $\sim40$ km
s$^{-1}$.

\subsubsection{Diffuse Interstellar Bands and the $E(B-V)$}

The centroid wavelengths and equivalent widths of 68 Diffuse Interstellar
Bands (DIBs) were measured. These DIBs are found over the entire spectral
coverage, although few are found at wavelengths longer than 7400 \AA. The DIBs
were identified using Jenniskens \& Desert (1994), from which also the
conversion factors between equivalent width and colour excess $E(B-V)$ for
each DIB were adopted. The derived colour excess is $E(B-V)=0.9^{+0.2}_{-0.1}$
mag, confirming the estimated $E(B-V)=1.0\pm0.2$ mag from a smaller selection
of DIBs by Molster et al.\ (1999) to whom we refer for a thorough discussion
on the distance to AFGL 4106. The heliocentric velocity is $v_{\rm
DIB}=+3\pm2$ km s$^{-1}$. The velocity of the DIBs differs significantly from
both the velocity of AFGL 4106 as measured from the spectra as well as the
system velocity of AFGL4106, confirming that the DIBs are of interstellar not
circumstellar origin.

\subsubsection{[N~{\sc ii}] emission lines}

%
%
\begin{figure}[tb]
\centerline{\psfig{figure=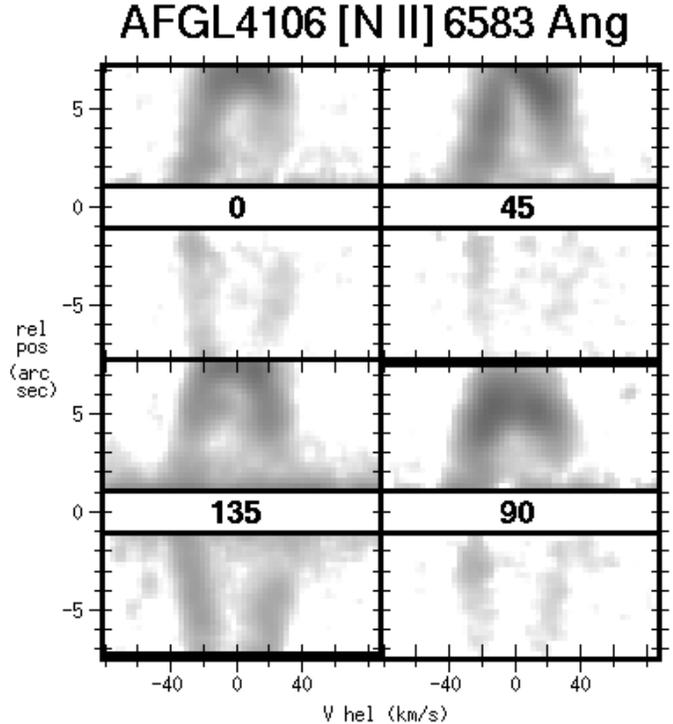,width=88mm}}
\caption[]{Spatial and kinematic map of [N~{\sc ii}] $\lambda$6583. The
stellar continuum is subtracted, and the flux scale is logarithmic. The
relative positions are defined as running from E to W at a position angle
0$^{\circ}$, SE to NW at 45$^{\circ}$, S to N at 90$^{\circ}$ and SW to NE at
135$^{\circ}$. The wavelength axes are transformed into heliocentric Doppler
velocity axes.}
\end{figure}

The [N~{\sc ii}] $\lambda$6583 line is the strongest emission line in AFGL
4106, and the best tracer of the kinematics of the circumstellar nebula.
Unlike the H$\alpha$, there is no stellar contribution to the line profile.
The underlying stellar continuum is subtracted by scaling the spatial profile
of the stellar spectrum outside the kinematic extend of the [N~{\sc ii}]
emission. In this way a map is constructed of the [N~{\sc ii}] emission as a
function of the position along the slit relative to the stellar position, and
the heliocentric Doppler velocity, for each of the four slit orientations
(Fig.\ 4). The flux scale is logarithmic, and the relative positions are
counted from E to W (at 0$^{\circ}$), from SE to NW (at 45$^{\circ}$), from S
to N (at 90$^{\circ}$) and from SW to NE (at 135$^{\circ}$). Close to the
stellar position the stellar continuum was too dominant over the [N~{\sc ii}]
emission to reliably correct for it, and this region in the four maps is
covered by the labels with the slit position angles.

The maps of the [N~{\sc ii}] emission are consistent with emission from a
radially expanding shell, with an expansion velocity of $\sim30$ km s$^{-1}$
and a radius of $\sim7^{\prime\prime}$. As in the H$\alpha$ emission map, the
NW side is brighter than the SE side. The expansion velocity is smaller than
the 40 km s$^{-1}$ derived from the Si~{\sc ii} absorption and also smaller
than the $v_{\rm exp}=35\pm1$ km s$^{-1}$ that we estimate from the $^{12}$CO
J=1$\rightarrow$0 emission at 115 GHz detected by Josselin et al.\ (1998).

%
%
\begin{figure}[tb]
\centerline{\psfig{figure=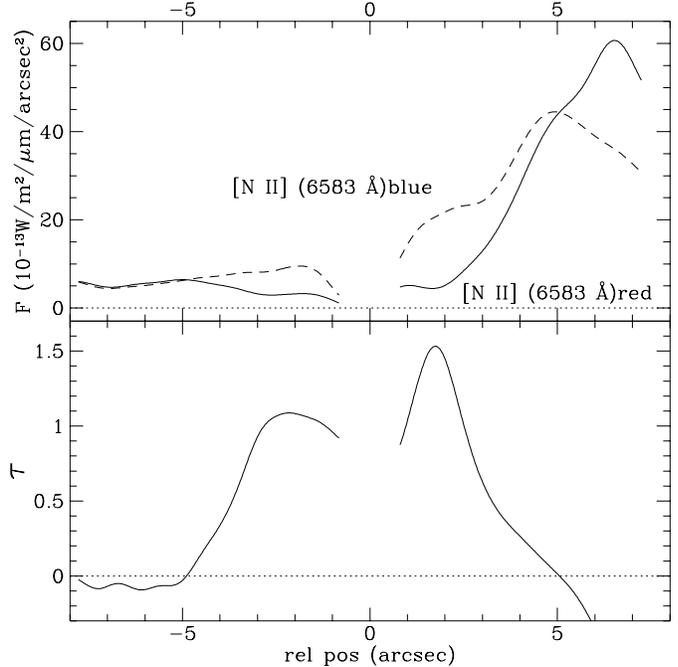,width=88mm}}
\caption[]{Spatial profile of [N~{\sc ii}] $\lambda$6583 (top panel), obtained
by summing the spectra taken at the four different position angles, with the
relative position defined as in Fig.\ 4. Positive relative positions roughly
correspond to NW. The blue- and red-shifted emission is shown individually.
The logarithm of their ratio yields the internal optical depth $\tau$ (bottom
panel) under the assumption that the [N~{\sc ii}] emission arises from a
spherically symmetric expanding shell.}
\end{figure}

The spectra at the four different position angles were summed and integrated
over the blue- and red-shifted velocity range, respectively. This results in
an effective cross section of the spatially extended emission complex running
roughly from SE to NW, i.e.\ from the minimum to the maximum of the emission
measure. The result is plotted in the top panel of Fig.\ 5. Assuming that the
source of the [N~{\sc ii}] emission is a spherically symmetric expanding
shell, the logarithm of the ratio of the blue- over the red-shifted emission
corresponds to the optical depth within the emitting shell. This is because
the red-shifted emission corresponds to the far side of the shell and has to
traverse through the dusty CSE within the shell before it emerges at the
observer's side which is also where the blue-shifted emission arises from. The
derived spatial profile of the optical depth is shown in the bottom panel of
Fig.\ 5.

The optical depth profile traces the dusty CSE out to a radius of
$5^{\prime\prime}$, which corresponds to the spatial extent of the 10 $\mu$m
emission (Molster et al.\ 1999), to the inner boundary of the H$\alpha$
maximum (Fig.\ 1) and the forbidden line maxima (Fig.\ 4). The optical depth
at $\lambda=6583$ \AA\ exceeds unity at a distance of $\sim2^{\prime\prime}$
from the star, but is lower closer to the star. This corresponds to the inner
cavity also seen in the 10 $\mu$m emission. The red-shifted [N~{\sc ii}]
emission becomes stronger than the blue-shifted emission at relative positions
${\gsim}5^{\prime\prime}$. This cannot be explained by optical depth effects
and must therefore indicate deviations from spherical symmetry in the density
and/or excitation conditions in that part of the nebula.

\subsubsection{[S~{\sc ii}] emission lines}

%
%
\begin{figure}[tb]
\centerline{\psfig{figure=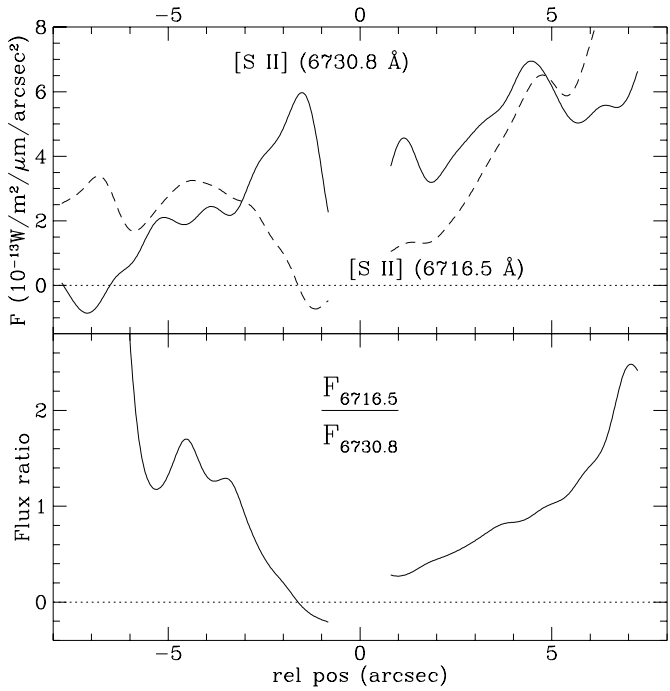,width=88mm}}
\caption[]{Spatial profile of [S~{\sc ii}] $\lambda$6717 and $\lambda$6731
(top panel), obtained as in Fig.\ 5 but integrated over the entire kinematic
extent of the emission line. Their flux ratio is plotted in the bottom panel.}
\end{figure}

The [S~{\sc ii}] $\lambda\lambda$6717,6731 emission is co-spatial with the
[N~{\sc ii}] emission, as seen from the spectra at the four different position
angles. The spectra are summed in the same way as in preparing the [N~{\sc
ii}] spatial profile in Fig.\ 5, but now the emission line is integrated over
its entire kinematic extent. This results in one-dimensional spatial emission
profiles for both components of the [S~{\sc ii}] doublet (Fig.\ 6, top panel).
The emission peaks in the NW (positive relative position), where the H$\alpha$
is brightest too.

The ratio of the emission profiles of the 6717 and 6731 \AA\ components (Fig.\
6, bottom panel) may be used to derive the local electron density $n_e$
(Osterbrock 1988). For $n_e<100$ cm$^{-3}$ this ratio is $>1.3$, with a
maximum of 1.42. For $n_e>10^4$ cm$^{-3}$ the ratio is $<0.5$, with a minimum
of 0.44. This holds for an electron temperature of $T_e=10^4$ K. The derived
electron density scales with $\sqrt{T_e}$. The CSE of AFGL 4106 outside of the
region of 10 $\mu$m emission has $n_e<100$ cm$^{-3}$, whereas the inner CSE
corresponding to the cavity in the 10 $\mu$m emission has $n_e>10^4$
cm$^{-3}$.

\subsubsection{H$\alpha$ emission line}

%
%
\begin{figure}[tb]
\centerline{\psfig{figure=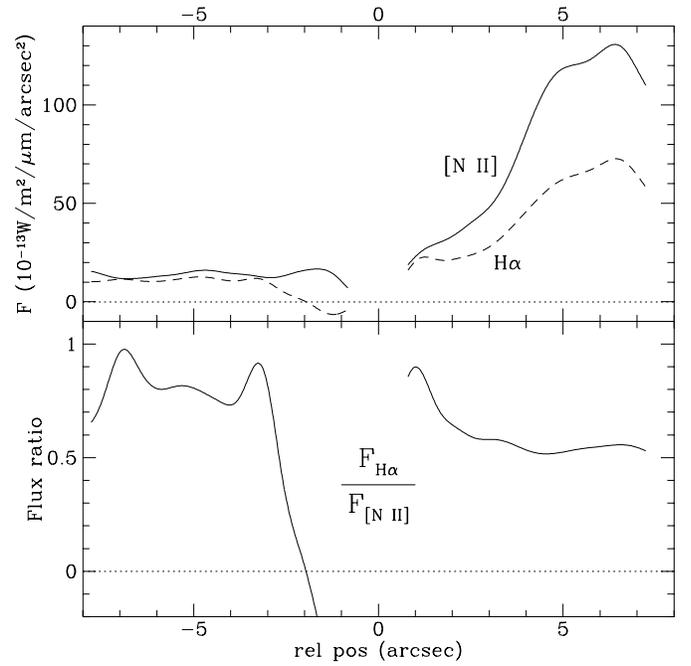,width=88mm}}
\caption[]{Spatial profiles of H$\alpha$ and [N~{\sc ii}]
$\lambda\lambda$6548,6583, obtained as in Fig.\ 5 (top panel). Their flux
ratio is plotted in the bottom panel.}
\end{figure}

%
%
\begin{figure}[tb]
\centerline{\psfig{figure=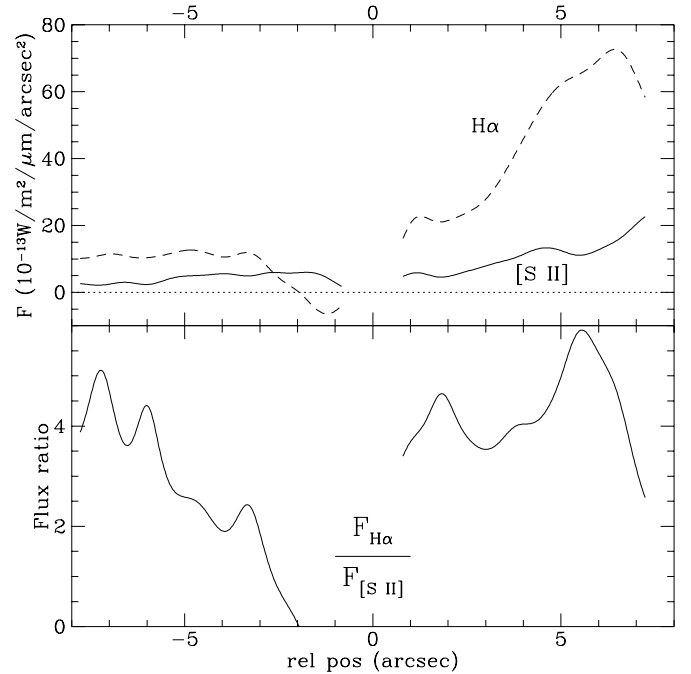,width=88mm}}
\caption[]{Spatial profiles of H$\alpha$ and [S~{\sc ii}]
$\lambda\lambda$6717,6731, obtained as in Fig.\ 5 (top panel). Their flux
ratio is plotted in the bottom panel.}
\end{figure}

Spatial profiles of the H$\alpha$ emission were created in the same way as for
[S~{\sc ii}]. The H$\alpha$ profile is compared with the profiles of [N~{\sc
ii}] (Fig.\ 7) and [S~{\sc ii}] (Fig.\ 8).

The flux ratios of H$\alpha$ and forbidden lines are in the range
$\log(F_{{\rm H}\alpha}/F_{\rm [N\ II]})\sim-0.3$ to $-0.05$ and
$\log(F_{{\rm H}\alpha}/F_{\rm [S\ II]})\sim0.3$ to $0.7$. This strongly
suggests excitation conditions similar to those in Planetary Nebulae but
different from those in H~{\sc ii} regions (Garc\'{\i}a-Lario et al.\ 1991).
The H$\alpha$ emission is co-spatial with the forbidden line emission. The
H$\alpha$/[N~{\sc ii}] ratio seems smallest in the brightest part of the
nebula around a relative position of $+5^{\prime\prime}$. The
H$\alpha$/[S~{\sc ii}] ratio seems to be smallest close to the star. This
effect may be caused by the spatial variation of the ratio of the [S~{\sc ii}]
components, with the $\lambda$6717 component essentially following the
H$\alpha$ emission.

\subsubsection{Fe~{\sc i} and [Ca~{\sc ii}] emission lines}

Though reported by Garc\'{\i}a-Lario et al.\ (1994), 16.6 \AA\ wide Fe~{\sc i}
emission centred at 6380.7 \AA\ could not be confirmed.

%
%
\begin{figure}[tb]
\centerline{\psfig{figure=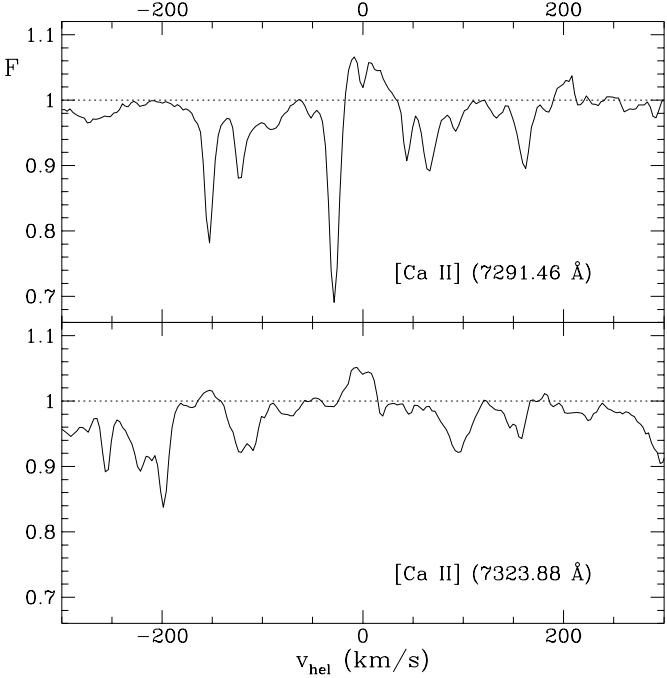,width=88mm}}
\caption[]{Spectrum around [Ca~{\sc ii}] $\lambda\lambda$7291,7324. The
wavelength axes are transformed into heliocentric Doppler velocity axes.}
\end{figure}

Instead, weak fluorescence emission of the 1F doublet of [Ca~{\sc ii}]
$\lambda\lambda$7291,7324 is detected (Fig.\ 9). The spectra have been
normalised in the same way as the Si~{\sc ii} spectrum (Fig.\ 3). The emission
is centred on the star and spatially unresolved. The line profiles are
distorted due to the presence of numerous telluric absorption lines but appear
to be kinematically symmetric with respect to the stellar velocity. The
emission extends between about $\pm20$ km s$^{-1}$, which is somewhat narrower
than the H$\alpha$, [N~{\sc ii}] and [S~{\sc ii}] emission. The equivalent
widths of the lines are $W_{7291}=-26\pm3$ and $W_{7324}=-31\pm4$ m\AA,
respectively. The line profiles peak at a flux density (above the stellar
continuum) of $F_{7291}\sim3.0\times10^{-12}$ and
$F_{7324}\sim2.3\times10^{-12}$ W m$^{-2} \mu$m$^{-1}$, respectively.

The location of the [Ca~{\sc ii}] emission can be further constrained by
estimating where the outflow reaches the critical density for the [Ca~{\sc
ii}] lines of $n_{\rm crit}=10^7$ cm$^{-3}$. However, this requires
calculation of the mass-loss rate. The mass-loss rate may be estimated from
the CO emission reported by Josselin et al.\ (1998). For the moment it is
assumed that this mass-loss rate is valid for the unresolved inner regions of
the nebula.

With distance $D$, outflow velocity $v$, CO abundance $f$ with respect to
H$_2$, CO peak antenna temperature $T$, and a simplified model for the CSE
(Kastner 1992), the total mass-loss rate can be derived:
\begin{eqnarray}
\dot{M} = 3.8\times10^{-12} & \hspace{-27mm} \left( \frac{T}{0.08}
\right)^\gamma \times \nonumber \\ & \times \frac{D^2\left[{\rm kpc}\right]^2
v^2\left[{\rm km s}^{-1}\right]^2}{f} {\rm M}_\odot {\rm yr}^{-1}
\end{eqnarray}
with the parameter $\gamma$ approximated as follows:
\begin{equation}
\gamma = \left\{ \begin{array}{ll} \frac{1}{2} & \mbox{for $\dot{M}<10^{-7}$
M$_\odot$ yr$^{-1}$} \\ \frac{2}{3} & \mbox{for $10^{-7}<\dot{M}<10^{-6}$
M$_\odot$ yr$^{-1}$} \\ \frac{3}{4} & \mbox{for $10^{-6}<\dot{M}<10^{-5}$
M$_\odot$ yr$^{-1}$} \end{array} \right.
\end{equation}
For higher mass-loss rates, more detailed modelling is required. In that case
the CO line profile will become saturated, and hence it is no longer the peak
antenna temperature, but the shape of the line profile that yields an
estimate of the mass-loss rate. With $v=35$ km s$^{-1}$, $f=2\times10^{-4}$,
$T=0.14$ K, and $\gamma=\frac{3}{4}$, the mass-loss rate is obtained:
\begin{equation}
\dot{M} = 3.5\times10^{-5} D^2\left[{\rm kpc}\right]^2 {\rm M}_\odot {\rm
yr}^{-1}
\end{equation}
This implies a very high mass-loss rate if the star were to be considerably
more distant than one kpc, but still an order of magnitude less than the
mass-loss rate derived by Molster et al.\ (1999) who estimate $\dot{M} \sim
3\times10^{-4} D\left[{\rm kpc}\right]$ M$_\odot$ yr$^{-1}$.

The angular separation $d$ from the star from which the [Ca~{\sc ii}] emission
arises can now be estimated assuming a constant outflow velocity of $v=35$ km
s$^{-1}$, and an average particle mass of $2.7\times10^{-24}$ g:
\begin{equation}
d = 1.4\times10^{13} D\left[{\rm kpc}\right] {\rm m} = 0.09^{\prime\prime}
\end{equation}
Hence for distances of $D{\lsim}20$ kpc the [Ca~{\sc ii}] emission must arise
from well within the inner cavity that is seen in the 10 $\mu$m emission and
in the internal extinction of the [N~{\sc ii}] emission. If the mass-loss rate
in those inner regions is lower than the CO estimate, then the [Ca~{\sc ii}]
emission is located even closer to the star. This is consistent with the
[Ca~{\sc ii}] emission being spatially unresolved.

\section{Modelling the circumstellar dust envelope}

The dust around AFGL 4106 causes extinction of line emission that originates
from behind the dusty CSE, and emission in the mid- and far-IR. Our spatially
resolved spectra of the nebular line emission were used to construct a spatial
profile of the optical depth of the dusty CSE, and the spatially resolved dust
emission at 10 $\mu$m was published by Molster et al.\ (1999). Here we make an
attempt to simultaneously model the spatial profiles of the optical depth and
of the dust emission.

\subsection{Circumstellar dust extinction}

Suppose that the bulk of the H$\alpha$ and [N {\sc ii}] emission originates
from a region just outside the dusty envelope. The emission from the front and
backside is observed at blue- and red-shifted velocities, respectively.
Assuming equally strong front and backside emission, the ratio of the observed
line flux at red- and at blue-shifted velocities directly yields the
differential extinction experienced by the backside emission after traversing
the dusty CSE. The [N~{\sc ii}] $\lambda$6583 line is best suited for this
method because it is strong and its intrinsic width is small compared to the
much larger thermal width of the H$\alpha$ line emission.

%
%
\begin{figure}[tb]
\centerline{\psfig{figure=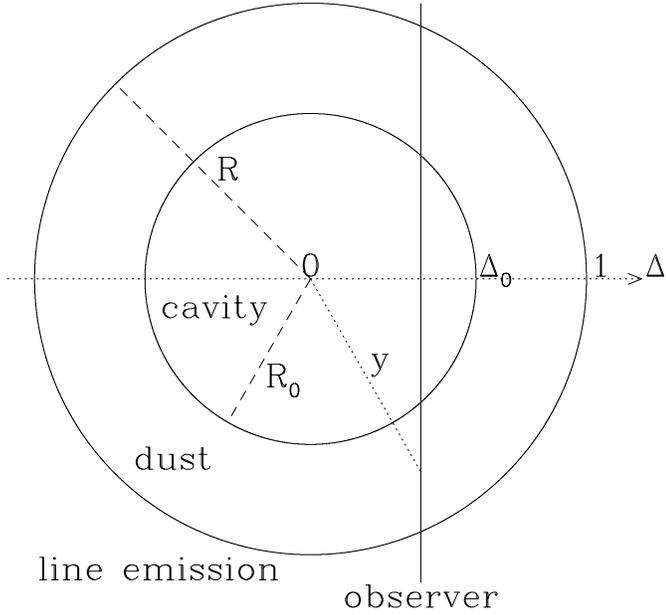,width=88mm}}
\caption[]{Schematic representation of the dust shell around AFGL 4106, and
the coordinates used in the modelling.}
\end{figure}

The mass-loss rate history through the dusty CSE is parameterised as:
\begin{equation}
\dot{M_r} = \dot{M_R} \left( \frac{r}{R} \right)^\alpha
\end{equation}
The case of $\alpha=0$ corresponds to a steady mass loss. The outflow velocity
through the dusty CSE is assumed to be constant. The optical depth
$\tau_{0.66}$ at 0.66 $\mu$m as a function of dimensionless projected distance
$\Delta$ is:
\begin{equation}
\tau_{0.66} = \zeta g(\Delta)
\end{equation}
where $\Delta=1$ at the outer radius $R$ of the dusty CSE, and
$\Delta=\Delta_0$ at the inner radius $R_0$ of the dusty CSE (Fig.\ 10). The
spatial profile of the observed optical depth has a shape:
\begin{equation}
g(\Delta) = \int_x^1 \frac{y^{\alpha-1}}{\sqrt{y^2-\Delta^2}} {\rm d}y
\end{equation}
with lower integration boundary:
\begin{equation}
x = \left\{ \begin{array}{ll} 1 & \mbox{if $\Delta>1$} \\ \Delta & \mbox{if
$\Delta_0<\Delta<1$} \\ \Delta_0 & \mbox{if $\Delta<\Delta_0$}
\end{array} \right.
\end{equation}
and a scaling parameter:
\begin{equation}
\zeta = \frac{3 \psi \dot{M_R}}{8 \pi R v s} \times \frac{Q_{0.66}}{a}
\end{equation}
with dust-to-gas ratio $\psi=5\times10^{-3}$, constant outflow velocity $v$,
dust grain specific mass $s=3$ g$\,$cm$^{-3}$ and radius $a$, and extinction
coefficient $Q_{0.66}=14 a\left[{\mu}{\rm m}\right]$ for oxygen-rich dust
(Volk \& Kwok 1988).

%
%
\begin{figure}[tb]
\centerline{\psfig{figure=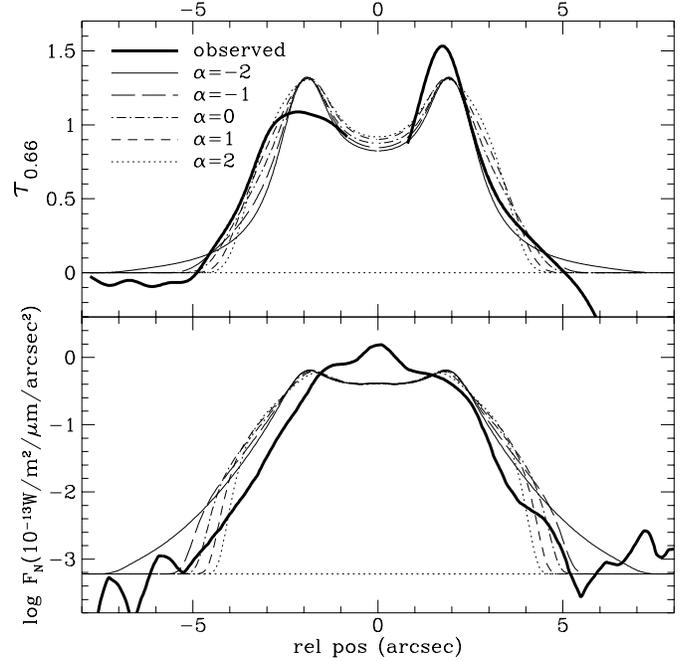,width=88mm}}
\caption[]{Top panel: spatial profile of the optical depth $\tau$ at 0.66
$\mu$m as derived from the [N~{\sc ii}] $\lambda$6583 emission. Bottom panel:
spatial profile of the 10 $\mu$m emission, for $p=1$. In both panels,
calculated profiles are plotted for different mass-loss rate histories
($\alpha$, see text).}
\end{figure}

For $\alpha\in\left\{-2,-1,0,1,2\right\}$ the function $g$ can be evaluated
analytically. To obtain a reasonable fit to the observed profile, the
parameters $\Delta_0$, $R$ and $\zeta$ need to be tuned accordingly. The model
profiles are convolved with a gaussian of $\sigma=0.36^{\prime\prime}$,
corresponding to the Full Width at Half Maximum of the stellar continuum in
the spatial direction ($0.85^{\prime\prime}$). We fitted by eye, aiming at
equal maximum optical depth for all $\alpha$. The results are plotted in Fig.\
11 (top panel) and summarised in Table 1.

\subsection{Circumstellar dust emission}

The spatial profile of the observed N-band flux density per square arcsecond
can be calculated as follows:
\begin{equation}
F_N = \zeta^\prime g^\prime(\Delta)
\end{equation}
with a shape:
\begin{equation}
g^\prime(\Delta) = \int_x^1 \frac{y^{\alpha-1}}{\sqrt{y^2-\Delta^2}} f(y)
{\rm d}y
\end{equation}
and a scaling factor:
\begin{equation}
\zeta^\prime = \left( \frac{1}{648,000} \right)^2
\frac{2 h c^2}{\lambda_N^5} \frac{Q_N}{Q_{0.66}} \zeta
\end{equation}
The wavelength dependence of the extinction coefficient $Q$ for oxygen-rich
dust (Volk \& Kwok 1988) is convolved with the TIMMI N-band filter curve,
yielding $Q_N=3.18 a\left[{\mu}{\rm m}\right]$. The dust emission is given by
a blackbody and experiences extinction on its way out of the CSE:
\begin{eqnarray}
f(y) = & \left[ e^{-\tau_N(y)} + e^{-\tau_N+\tau_N(y)} \right] \times
\nonumber \\ & \times \left[ \exp \left( \frac{h c}{\lambda_N k T(y)} \right) -
1 \right]^{-1}
\end{eqnarray}
with total optical depth $\tau_N$ along line-of-sight $\Delta$:
\begin{equation}
\tau_N = \frac{Q_N}{Q_{0.66}} \tau_{0.66}
\end{equation}
and $\tau_N(y)$ the partial optical depth through the CSE between us and the
point $y$. The dust is assumed to be in radiative equilibrium with the diluted
and attenuated stellar radiation field. The stellar radiation is generated by
two stars that have temperatures $T_\star$ and $\phi_TT_\star$, and
luminosities $L_\star$ and $\phi_LL_\star$. We adopt $Q\propto\nu^p$ for the
overall frequency dependence of the extinction coefficient, with $p$ between
$\sim1$ and 2. The temperature profile is then given by (see also Sopka et
al.\ 1985):
\begin{eqnarray}
T(y) = \xi_1 \left[ \frac{1}{y^2} \int_0^\infty \right. & \gamma(\eta) \,
\eta^{3+p} \left( \exp \left( \frac{\eta}{T_\star} \right) - 1 \right)^{-1}
\times \nonumber \\ & \left. \times \exp \left(\xi_2 \eta^p
\int_{y}^{\Delta_0} \delta^{\alpha-2} {\rm d}\delta \right) {\rm d}\eta
\right]^\omega
\end{eqnarray}
with $\omega=1/(4+p)$ and
\begin{equation}
\xi_1 = \left[ \frac{\Delta_\star^2}{4(3+p){\rm !} Z(4+p)} \right]^\omega
\end{equation}
where the values of the Euler Zeta function $Z(x)=\sum_{n=1}^\infty n^{-x}$
for $p=1$ and $p=2$ are $Z(5)\sim1.037$ and $Z(6)\sim1.0173$, respectively,
and
\begin{equation}
\xi_2 = \frac{\zeta}{2 \pi} \left( \frac{\lambda_{0.66} k}{h c} \right)^p
\end{equation}
The correction factor $\gamma$ is:
\begin{equation}
\gamma(\eta) = 1 + \frac{\phi_L}{\phi_T^4} \left[ \frac{\exp \left( \eta /
T_\star \right) - 1}{\exp \left( \eta / \left( \phi_TT_\star \right) \right) -
1} \right]
\end{equation}
where $\gamma=1$ in the case of a single star and $\gamma=2$ in the case of
two identical stars.

Model profiles are calculated for the mass-loss histories examined above. The
profiles are convolved with a gaussian of $\sigma=0.26^{\prime\prime}$, which
corresponds to the diffraction limit of the ESO 3.6 m at 10.1 $\mu$m. We adopt
a stellar temperature $T_\star=7500$ K and $\phi_T=4000/7500$, with a
luminosity ratio of $\phi_L=1/1.8$ (Molster et al.\ 1999, van Winckel et
al.\ 1999). The only free parameter left is the stellar radius $R_\star$ of
the primary. We fitted by eye, aiming at equal emission in the centre for all
$\alpha$. The results for $p=1$ are plotted in Fig.\ 11 (bottom panel) and
summarised in Table 1.

\subsection{Model results}

%
%
\begin{table*}[tb]
\caption[]{Modelling parameters of the spatial profiles of the optical depth
at 0.66 $\mu$m and the 10 $\mu$m flux density. Five models are concerned, with
$p=1$ and different mass-loss rate histories parameterised by $\alpha$.
Distance dependent quantities are given assuming a distance $D=1$ kpc.}
\begin{tabular}{lrlllllllllll}
\hline\hline
\#                                &
$\alpha$                          &
$R$[$^{\prime\prime}$]            &
$t$[yr]                           &
$R_0$[$^{\prime\prime}$]          &
$t_0$[yr]                         &
$R_\star$[$^{\prime\prime}$]      &
$R_\star$[R$_\odot$]              &
$\zeta$                           &
$\dot{M}$[M$_\odot\,$yr$^{-1}$]   &
$\dot{M}_0$[M$_\odot\,$yr$^{-1}$] &
$M$[M$_\odot$]                    &
$L_{\rm tot}$[L$_\odot$]          \\
\hline
A & \llap{$-$}2 & 7   & 948 & 1.9  & 257 & 3.4$\times 10^{-4}$ & 73 &
0.05 & $1.0\times10^{-5}$ & $1.4\times10^{-4}$ & 0.027 & 1.5$\times 10^4$ \\
B & \llap{$-$}1 & 4.9 & 664 & 1.85 & 251 & 3.1$\times 10^{-4}$ & 67 &
0.28 & $4.1\times10^{-5}$ & $1.1\times10^{-4}$ & 0.026 & 1.3$\times 10^4$ \\
C &           0 & 4.4 & 596 & 1.8  & 244 & 3.0$\times 10^{-4}$ & 64 &
0.60 & $7.9\times10^{-5}$ & $7.9\times10^{-5}$ & 0.028 & 1.2$\times 10^4$ \\
D & \llap{$+$}1 & 4   & 542 & 1.7  & 230 & 2.9$\times 10^{-4}$ & 62 &
1.05 & $1.3\times10^{-4}$ & $5.3\times10^{-5}$ & 0.028 & 1.1$\times 10^4$ \\
E & \llap{$+$}2 & 3.7 & 501 & 1.6  & 217 & 2.8$\times 10^{-4}$ & 60 &
1.60 & $1.8\times10^{-4}$ & $3.3\times10^{-5}$ & 0.027 & 1.0$\times 10^4$ \\
\hline
\end{tabular}
\end{table*}

From the output parameters $R$, $R_0$, $R_\star$ and $\zeta$ the mass-loss
rates at the inner and outer radii ($\dot{M}_0$ and $\dot{M}$), the total mass
$M_{\rm CSE}$ in the dusty CSE and the total luminosity $L$ of the object are
derived, assuming a distance $D=1$ kpc. The results for $p=1$ are summarised
in Table 1. The emission profiles for $p=2$ are indistinguishable from those
for $p=1$, with the only difference being the stellar radius (and hence the
stellar luminosity) which is almost six times larger for $p=1$. The CSE radii
are related to the time in the past when the material at those radii had been
expelled, assuming a constant expansion velocity of $v_{\rm exp}=35$ km
s$^{-1}$. The stellar radius of the hot star is expressed both in arcseconds
and in solar radii. Again, these conversions are made under the assumption of
a distance $D=1$ kpc. Radii (not angular), dynamical times, and mass-loss
rates scale linearly with distance, whereas the CSE mass and the (total)
luminosity scale quadratically with distance.

The mass-loss history (or $alpha$) is not much constrained by the optical
depth and the 10 $\mu$m emission. The dynamical times do not depend strongly
on the mass-loss history, and we find almost identical total CSE masses of
$M=0.027$ M$_\odot D^2$[kpc]$^2$. Our model does not reproduce the 10 $\mu$m
emission exactly. Apart from an observed point-source component in the centre
with a flux density of $80\times10^{-13}$ W m$^{-2} \mu$m$^{-1}$ ($\equiv90$
Jy) that is not reproduced by our model, the observed emission is more compact
than our model produces. Lowering the effective temperature of the star(s)
and/or including absorption of stellar light by a fixed column density before
the stellar radiation reaches the spatially resolved dust shell both lead to a
larger derived stellar radius and luminosity --- and thus reduces the need for
an additional point source in the centre --- but the shape of the emission
profiles remains virtually identical.

\section{Discussion}

\subsection{Binary orbit}

The preliminary binary velocity curve that we present can only be used as an
indication for the maximum masses of the stellar components of the binary in
AFGL 4106. Their current masses are limited to $\lsim15\sin^{-3}i$ M$_\odot$,
which is consistent both with the masses and luminosities derived for a
distance of 3.3 kpc by Molster et al.\ (1999) and with the high ratio of 60
$\mu$m flux density over CO brightness temperature (see Josselin et al.\
1996) if the inclination angle $i$ is close to $90^\circ$ (edge-on).

\subsection{Structure of the ionized nebula}

Our coronographic imaging and spectroscopy provide new insight into the
complex structure of the circumstellar envelope of AFGL 4106. Emission from
H$\alpha$ and the forbidden lines [N~{\sc ii}] and [S~{\sc ii}] originates
mainly from a thin shell envelopping a roughly spherical thicker shell of
dust. This dust shell extends from $\sim2.7\times10^{16} D$[kpc] cm to
$\sim6.6\times10^{16} D$[kpc] cm and has an optical depth around H$\alpha$
close to unity. There is evidence for part of the nebular emission to also
arise from within the dust shell: the electron density as traced by the ratio
of the two components of the [S~{\sc ii}] doublet is higher at smaller
projected distances from the star. A mixed dust/ion shell may result from a
clumpy dust distribution, with some of the stellar UV radiation able to
permeate the dust shell. Indeed the 10 $\mu$m image of AFGL 4106 (Molster et
al.\ 1999) suggests a clumpy density or temperature distribution. An
extinction map such as H$\alpha$/H$\beta$ would be useful to trace the dust,
whereas a radio map obtained in the free-free continuum may be used to trace
the distribution of ionized matter. Density-bounded ionized nebulae like the
one around AFGL 4106 are rare and presumably short-lived, but not unique:
similar nebulae are observed around Luminous Blue Variables (e.g\ AG Car) and
Wolf-Rayet stars (Voors et al., in preparation).

Unless a significant population of very small grains is postulated, dust
temperatures in the detached dust shell around AFGL 4106 do not reach more
than a few $\times10^2$ K and hence thermal emission from dust grains cannot
explain the part of the near-IR emission observed by Garc\'{\i}a-Lario et al.\
(1994) that is (projected) co-spatial with the dust shell. Thus it seems more
likely that the near-IR emission is due to either free-free emission from the
ionized component in the dust shell or starlight that is scattered by the
large grains ($a\sim1 \mu$m) found by Molster et al.\ (1999). The line
intensity ratios of H$\alpha$, [N~{\sc ii}] and [S~{\sc ii}] unambiguously
point at excitation conditions similar to those found in Planetary Nebulae
(see Garc\'{\i}a-Lario et al.\ 1991). In the bow-shaped region where the line
emission reaches its maximum the excitation mechanism may have a shock
component, resulting in enhanced emission by [N~{\sc ii}] over that by
H$\alpha$. The shock may be due to the collision of the CSE of AFGL 4106 with
the interstellar medium if the system is moving with a supersonic velocity
with respect to the interstellar medium. H$\alpha$ emission from regions
outside of the bow shape (up to about twice the outer radius of the dust
shell) may be explained by an H~{\sc ii} region being formed.

\subsection{Evidence for on-going mass loss}

AFGL 4106 is experiencing on-going mass loss. Evidence for this is seen in
enhanced absorption in the short-wavelength wings of Si~{\sc ii}, Fe~{\sc
i,ii} and N~{\sc i} lines. Fluorescence emission of [Ca~{\sc ii}] (see also
Riera et al.\ 1995) originates from material that has been expelled not more
than a few decades ago. Furthermore, the core emission in the 10 $\mu$m image
must be of circumstellar origin (either dust or ionized gas). If dust is
present close to the star then it must be optically thin at visual
wavelengths. The reason for this is that our modelling of the 10 $\mu$m
emission yields a luminosity that is in perfect agreement with the integrated
bolometric luminosity and reasonable estimations for interstellar extinction
(Garc\'{\i}a-Lario et al.\ 1994; Molster et al.\ 1999) if the dust extinction
coefficient index $p=1$ (as in Volk \& Kwok 1988).

\subsection{Mass-loss history}

The mass contained in the dust shell is a few $\times10^{-2} D^2$[kpc]$^2$
M$_\odot$ and $\sim13$ times less than the mass as derived by Molster et al.\
(1999). This is in part due to the choices by Molster et al.\ for the
(smaller) inner and (larger) outer radii of the dust shell that together yield
a duration of the dusty mass loss $t-t_0$ that is about three times as long as
we estimate here. They assume a gas-to-dust ratio of 100 whereas here we adopt
a value of 200, but the time-averaged dust mass-loss rate that Molster et al.\
derive from their modelling of the spectral energy distribution is an order of
a magnitude higher than we estimate here. This may be the result of
differences in the optical constants of the dust species used in their
modelling using large grains ($a\sim1 \mu$m) from those of the dust in Volk \&
Kwok (1988) that we use here. The mass-loss rate estimate from the CO
emission, however, is in between the mass-loss rates that we estimate for the
inner and outer radii of the dusty CSE, and thus seems to favour our more
moderate mass-loss rates over the extremely high mass-loss rates estimated by
Molster et al.\ (1999). The discrepancy might be resolved by storing most of
the dust mass into large grains at large distances from the star, outside the
CSE as traced by the 10 $\mu$m emission, yielding the observed emission at
wavelengths longward of $\sim50 \mu$m.

The expansion velocity has increased over time from $\sim30$ km s$^{-1}$ at
the time when the material at the outer rim of the dust shell was expelled
(tracer: [N~{\sc ii}]) through $\sim35$ km s$^{-1}$ (tracer: CO) up to
$\sim40$ km s$^{-1}$ at present (tracers: Si~{\sc ii}, Fe~{\sc i,ii} and
N~{\sc i}). In combination with a steady mass loss this velocity evolution
could yield the radial density dependence of $\rho{\propto}r^{-2.2}$ as derived
by Molster et al.\ (1999). This consideration suggests that of our models,
model C is to be favoured: a constant mass-loss rate during the formation of
the dust shell.

\acknowledgements{We greatly appreciate having been granted Director's
Discretionary Time to obtain the NTT images and spectra. We would like to
thank the anonymous referee for her/his suggestions. This research was partly
supported by NWO under Pionier Grant 600-78-333. FJM acknowledges support from
NWO grant 781-71-052. Jacco n\~{a}o pode descrever o muito que agradece ao
anjo Joana.}

\end{document}